\documentclass{elsart1p}
\usepackage{graphicx}
\usepackage{epstopdf}
\usepackage{amssymb}
\usepackage{amsmath}
\usepackage{bm}

\newcommand{\Nc}{N_c}
\newcommand{\Tc}{T_c}

\newcommand{\Nf}{N_f}

\begin{document}

\begin{frontmatter}


% use the thanksref command within \title, \author or \address for
%footnotes;
% use the corauthref command within \author for corresponding author
%footnotes;

\title{Suppression of the shear viscosity near the critical temperature in
hot QCD}


\author[hidaka]{Yoshimasa Hidaka} and 
\author[pisarski]{Robert D. Pisarski}
\address[hidaka]{RIKEN BNL Research Center, Brookhaven National Laboratory,
Upton, NY 11973, USA}
\address[pisarski]{Department of Physics, Brookhaven National Laboratory,
Upton, NY 11973, USA}
\begin{abstract}
We consider QCD near but above critical temperature $\Tc$. The pressure,
susceptibilities and the
renormalized Polyakov loop --- 
which is an order parameter for the deconfining phase transition
--- dramatically change up to temperatures a few times $\Tc$. 
We refer to this region as a "semi"-QGP, 
where partial confinement plays important role. 
We show that the shear viscosity $\eta$ is suppressed by two
powers of the
Polyakov loop. This suggests that $\eta/T^3$ decreases markedly as QCD
cools down to temperatures near $\Tc$.
We also show a ratio of the viscosity to the entropy becomes small near $\Tc$
\cite{Hidaka:2008dr}.
\end{abstract}
%
%\begin{keyword}
%Quark gluon plasma \sep
%transport coefficient
%\PACS
%25.75.Nq \sep 12.38.Mh 
%\end{keyword}
\end{frontmatter}
\section{Introduction: Semi-QGP}
To understand nature of the deconfining phase transition in QCD
at finite temperature and/or density
is one of current topics of great interest in hadron physics.
The experimental results on heavy-ion collisions at relativistic heavy ion
collider (RHIC),
which may have probed a quark-gluon plasma (QGP), have brought remarkable
results.
One of them is elliptic flow suggesting a small ratio of the shear
viscosity to the entropy. 
This result cannot be reproduced by an ordinary weakly coupled plasma, 
so it has been described, rather, as a strongly coupled quark-gluon plasma.
In this talk, we discuss how the shear viscosity can be small even in
perturbation theory, for moderate values of the coupling 
near the phase transition.

At the deconfining phase transition, 
the physical degrees of freedom
change from mesons and baryons, for which the pressure is of order one,
to quarks and gluons, with
pressures of order $\Nf\Nc$ and $\Nc^{2}$, where $\Nc$ and $\Nf$
are the number of colors and flavors.
It is useful to view deconfinement as the ionization of color charge. 
In the confined phase, there is no ionization of color. 
Conversely, far into the deconfined phase, color is completely ionized.
The ionization in non Abelian gauge theory is characterized by the
renormalized Polyakov loop, which is an order parameter for 
a global $Z(\Nc)$ symmetry%
\footnote{The expectation value of the renormalized Polyakov loop has an
ambiguity associated with the 
renormalization scheme. Here we assume a scheme in which zero
point energy vanishes.}.
While the Polyakov loop is a strict order parameter in pure Yang-Mills
theory, in QCD it is only an approximate order parameter.
Lattice simulations, however, find that the renormalized Polyakov loop
is numerically close to a 
good order parameter, e.g.,  the susceptibility of the
Polyakov loop has a sharp peak at $\Tc$~\cite{loop2}.
At extremely high temperature, quarks and gluons are almost freely moving,
and one has a complete QGP,
where the renormalized Polyakov loop is  near one, and is insensitive to
temperature.
In the hadronic phase, the value of Polyakov loop is almost zero up to near
$\Tc$.

In the intermediate region, it is expected that the Polyakov loop changes
strongly with temperature, and represents the partial ionization of color.
We refer to this region as a "semi"-QGP.  Obvious questions are:
how wide is the semi-QGP, and how do observables change in it?
Lattice simulations provide partial answers to these questions:
for three colors,
they find that a semi-QGP window is $\Tc$ to $4\Tc$ in
pure Yang-Mills theory
and $0.8 \, \Tc$ to $\sim 2 -3\, \Tc$ with $2+1$ flavors of dynamical
quarks~\cite{loop2,loop1}.
The RHIC experiment could probe a semi-QGP.
We expect that not only static observables, such as the pressure,
but also dynamical quantities, such as transport coefficients,  
change strongly in the semi-QGP.

\section{Shear viscosity in the semi-QGP}
The transport coefficients are parameters in hydrodynamics which can not be
determined by itself.
They are calculated in several methods such as Kubo formula and kinetic
theory.
Here, we consider how the shear viscosity changes in the semi-QGP.
We do not give the explicit calculation but a simple explanation for how the
shear viscosity can be small even with a small coupling
constant in the semi-QGP.
In classical transport theory, the shear viscosity $\eta$ is given by
\begin{equation}
\eta\approx \frac{1}{3}n \bar{p} \lambda \;,
\end{equation}
where $n$ is the number density of quarks and gluons, $\bar{p}$ is the mean
momentum, and $\lambda$ is the 
mean free path. There are three possibilities to get small viscosity: the
number of particles, mean momentum,
or mean free path become small. In a relativistic plasma, $\bar{p}\sim T$
for light particles.
We assume that the mean momentum does not get small, so that we treat it as
of order $T$.
In general, the mean free path depends on the number of particles.
One may expect that the more dilute the system is, the longer the mean free
path is.
In fact, in classical transport theory,
the mean free path is $\lambda\sim 1/(n\sigma)$, where $\sigma$ is the
transport cross section;
thus, $\eta\sim T/\sigma$.
This is true in a QGP at very high temperatures,
where $n\sim \Nc^{2}T^{3}$, $\sigma \sim (g^{4}\ln 1/g)/T^{2}$,
and $g$ is the gauge coupling constant. The logarithmic term, $\ln 1/g$, is
known as the Coulomb logarithm, and is caused by 
an infrared singularity at the forward scattering region.
As a result, $\eta\sim T^{3}/(g^{4}\ln1/g)$, which gives the same result of
the Boltzmann equation, up to overall constant~\cite{amy}.

In the semi-QGP, the situation is different. 
The interaction and the number density of the particles may depend on their
color structures because of the partial ionization of color.
Then, the average mean free path is 
\begin{equation}
\lambda^{-1}\sim\frac{\sum_{{a,b}}  n_{a}n_{b} \sigma_{a,b} }{\sum_{a}  
n_{a} } \;.
\end{equation}
where $\sigma_{a,b}$ and $n_{a}$ is the transport cross section and number
density for color indices $a,b$.
If the transport cross section  is independent of color, the mean free path
reduces to $\lambda\sim 1/(n\sigma)$.

We found $\sum_{a} n_{a}  \sim \Nc^{2}\ell^{2}$,  and $\sum_{{a,b}}
n_{a}n_{b}\sigma_{a,b} \sim \Nc^{4}\ell^{2}\sigma$
in the semi-QGP \cite{Hidaka:2008dr}, where $\ell$ is the expectation value
of the Polyakov loop.
Therefore the mean free path $\lambda\sim \Nc^{2}/\sigma$ does not become
small as the density decreases;
the shear viscosity becomes small, $\eta\sim T\ell^{2}/\sigma$,  by two
powers of the Polyakov loop.
This implies that the interactions canceling color are more favored, i.e.,
the correlation of colored particles 
is stronger even for small values of the coupling constant.
Consequently, the shear viscosity can be small in the semi-QGP.

\begin{figure}
\begin{center}
  \includegraphics[scale=.77]{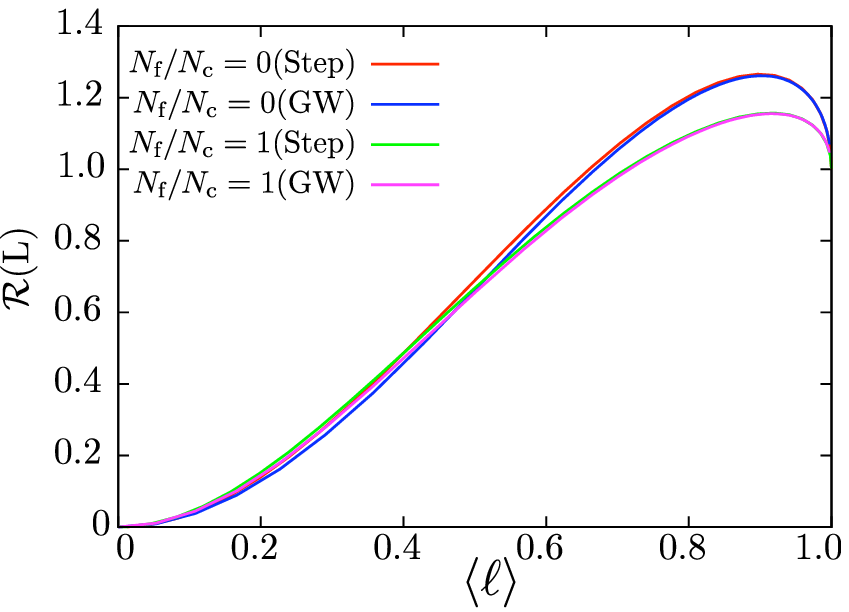} 
  \includegraphics[scale=0.2]{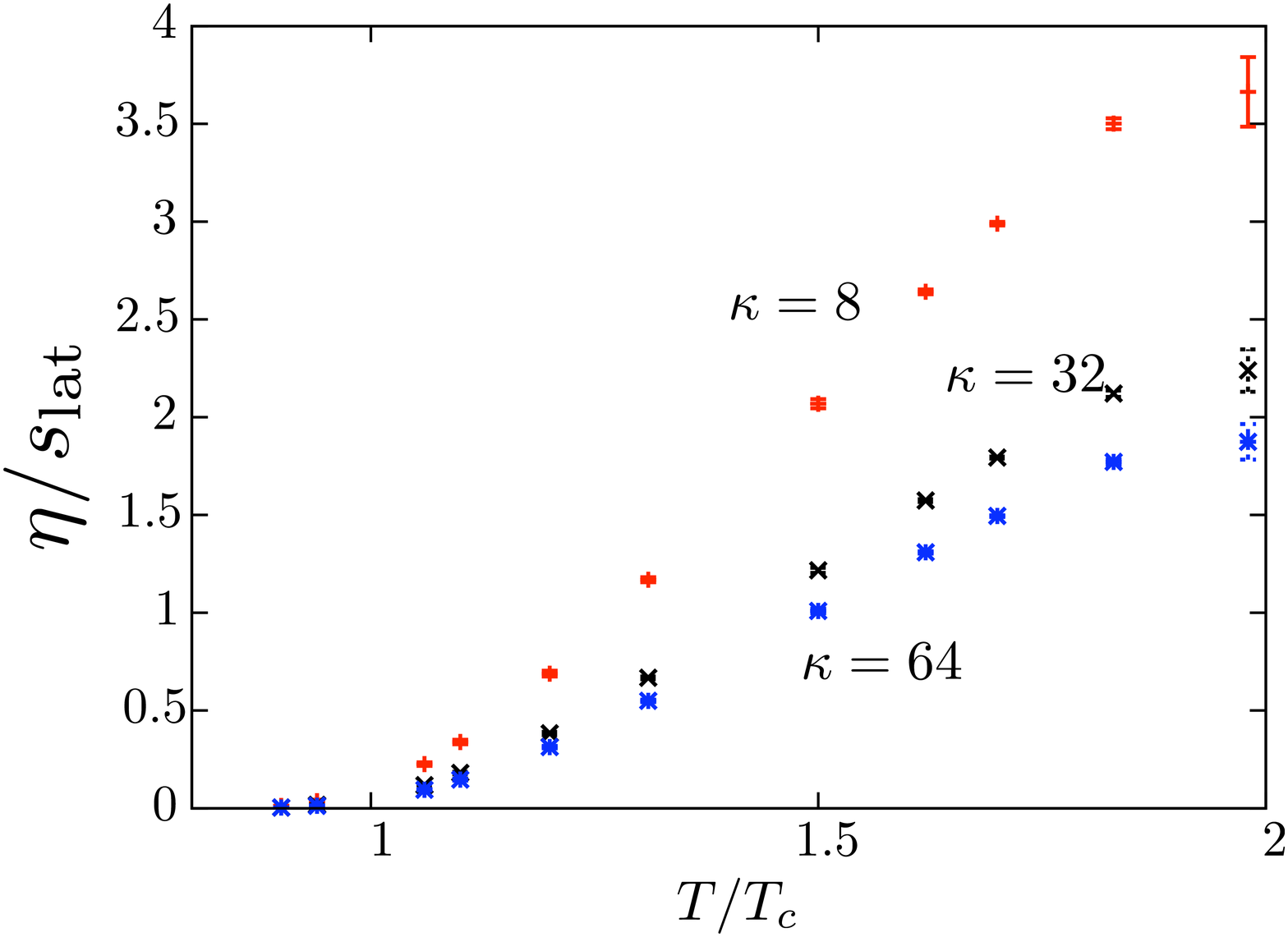} 
  \end{center}
\caption{
The function $\mathcal{R}(L)$ of Eq. (\ref{ratio}), versus $\ell$.
``Step'' and ``GW'' denote  eigenvalue distributions of the Wilson line
with a simple step-function
and  that in the Gross-Witten matrix model \cite{GW}, 
respectively (left).
The ratio of our viscosity to the entropy calculated on the lattice
(right).}
\label{fig}
\end{figure}

\section{Numerical Results}
In the previous section, we discussed how the shear viscosity is
suppressed for small values of the Polyakov loop.
In this section, we show numerical results. 
We make numerous drastic assumptions to characterize the semi-QGP.
We assume that the coupling constant is small even near $\Tc$.
For simplicity, we take 
the number of colors and flavors to be large.
We work in
a semiclassical approximation, so
that we can treat 
the Polyakov loop as a constant classical background field 
about which we expand.  The background field
corresponds to
an eigenvalue distribution of the Polyakov loop. 
Since this distribution is so far unknown, we need to make
assumptions about it.
We use two forms to check the sensitivity of our results.
The first is to take a simple step function, of width $\beta$, about
the origin. 
The other is to take a distribution as in the Gross-Witten matrix
model~\cite{GW}. 
We then calculate the shear viscosity by employing a Boltzmann equation 
to leading order in $g$ and $\log(1/g)$~\cite{Hidaka:2008dr}.
We take the following parameterization to compare with previous
work~\cite{amy}:
\begin{equation}
\mathcal{R}(L) = \frac{\eta(L)}{\eta(0)}  \,,
\hspace{1cm}
\eta(0) =\frac{c_\eta}{g^4\ln(\kappa/g)} \;,
\label{ratio}
\end{equation}
where $\eta(0)$ is the viscosity in absence of the background.
The coefficient $c_{\eta}$ is the numerical constant depending on the
number of colors and flavors.
$\kappa$ is higher order correction beyond the leading-log order; here we
treat it as a parameter.

In Fig.~\ref{fig}, we show  numerical results of $\mathcal{R}(L)$ with
$\Nf/\Nc=0,1$.
The differences between eigenvalue distributions are very small, most a few
percent over the entire range of $\ell$.
We find that the shear viscosity is suppressed near $T_c$ or at a small
$\ell$, as we discussed in the previous section.
A bump is observed near $\ell\simeq 0.9$. 
The value is 25\% larger than $\eta(0)$, which is analytically estimated
$\sim 1+1.47\sqrt{1-\ell}$.
We expect that this non-analytic behavior will be washed out by corrections
to higher order.

For hydrodynamics, a more important quantity than $\eta$ is 
the dimensionless ratio, $\eta/s$, where
$s$ is the entropy density.  This
is related to a diffusion constant, $D=\eta/(sT)$.
In Fig.~\ref{fig}, 
we compare the shear viscosity to the lattice entropy $s_\text{lat}$.
The coupling constant used is a one-loop running coupling with
$\alpha_{s}(\Tc)=1/3$;
an unknown parameter beyond the leading-log order is taken $\kappa=8$, $32$
and $64$.
The result strongly depends on the choice of $\kappa$; however, in any
case, 
$\eta/s_\text{lat}$ is small near $\Tc$, and becomes larger as temperature
increases. 
\section{Summary}
We have shown that the shear viscosity is suppressed by two powers of the
Polyakov loop in the semi-QGP region near $\Tc$.
It is natural to suspect that heavy ion collisions at RHIC have probed some
region in the semi-QGP.
Since one needs a small value of the shear viscosity to fit the
experimental data, perhaps one is near $\Tc$.  
Heavy ion collisions at the
LHC may probe temperatures which are significantly higher, possibly well
into the complete QGP.  If so, then at small times collisions at the LHC
creates a system with 
large shear viscosity; as the system cools through $\Tc$, the shear
viscosity then drops.
Thus the semi-QGP predicts that at short times, the hydrodynamic behavior
of
heavy ion collisions at the LHC is qualitatively
different from that at RHIC.

%%%%%%%%%%   Acknowledgments   %%%%%%%%%%
\ack{
This research was supported in part by the RIKEN BNL Research 
Center and by the U.S. Department of Energy under
Cooperative Research Agreement No. DE-AC02-98CH10886.
R.D.P. also thanks
the Alexander von Humboldt Foundation for their support.
}

\end{document}